\documentclass[preprint2]{aastex}
\received{Apr 13 2004}
\accepted{June 14 2005}
\slugcomment{}
\shorttitle{Black Hole Masses and Host Galaxy Evolution}
\shortauthors{Woo et al.}

\begin{document}

\title{Black Hole Masses and Host Galaxy Evolution of Radio-loud Active Galactic Nuclei}

\author{Jong-Hak Woo\altaffilmark{1,2},
C. Megan Urry\altaffilmark{3},
Roeland P. van der Marel\altaffilmark{4},
Paulina Lira\altaffilmark{5},
Jose Maza\altaffilmark{6}}

\altaffiltext{1}{Department of Astronomy and Yale Center for Astronomy and Astrophysics, Yale University, P.O. Box 208101, New Haven, CT 06520-8101; jhwoo@astro.yale.edu}
\altaffiltext{2}{Current address: Department of Physics, University of California at Santa
Barbara, CA 93106; woo@physics.ucsb.edu}
\altaffiltext{3}{Department of Physics and Yale Center for Astronomy and Astrophysics, Yale University, P.O. Box 208121, New Haven, CT 06520-8121; meg.urry@yale.edu}
\altaffiltext{4}{Space Telescope Science Institute, 3700 San Martin Dr. Baltimore MD 21218; marel@stsci.edu                      }
\altaffiltext{5}{Departamento de Astronom?a, Universidad de Chile, Casilla 36-D, Santiago, Chile; plira@das.uchile.cl}
\altaffiltext{6}{Departamento de Astronom?a, Universidad de Chile, Casilla 36-D, Santiago, Chile; jmaza@das.uchile.cl}

\begin{abstract}
We report stellar velocity dispersion measurements for a sample of 28 AGN
host galaxies including our previous work.
Using the mass-dispersion ($M_{\bullet}-\sigma$) and the fundamental plane relations,
we estimate the black hole mass for a sample of 66 BL Lac objects and investigate the
role of black hole mass in the energetics of BL Lac objects.
The black hole mass range for different BL Lac spectral types
%(i.e. X-ray to radio flux ratio)
is similar, $10^{7} < M_{\bullet} < 4 \times 10^{9}$.
Neither X-ray nor radio luminosity correlates with black hole mass.
Low-frequency-peaked BL Lac objects have higher Eddington ratios on average,
because of either more beaming or higher intrinsic power.
For the black hole mass range $3 \times 10^{7} <  M_{\bullet} < 10^{9}$,
the radio luminosity of BL Lac objects and flat-spectrum radio quasars
spans over 4 orders of magnitude,
with BL Lac objects being low-power AGNs.
We also investigate the evolution of host galaxies for 39 AGNs out to $z \approx 0.5$ with measured
stellar velocity dispersions.
Comparing the mass-to-light ratio evolution in the observed frame
%(without $K$-correction)
with population synthesis models,
we find that single burst star formation models with $z_{form} = 1.4^{+0.9} _{-0.2} $ are consistent with
the observations.    
%We provide $K$-correction values for a few stellar population models with a concordant cosmology.
From our $z_{form}=1.4$ model, we estimated
%After $K$-correction using our $z_{form}=1.4$ model, we estimated
the intrinsic mass-to-light ratio evolution in the Cousins $R$ band,
$\Delta log (M/L)/ \Delta z = -0.502 \pm 0.08$, consistent with that of normal early type galaxies.
%The host galaxie of our sample belong to the most massive galaxies
\end{abstract}


\section{Introduction}

The role of black holes in galaxy formation and evolution appears to be significant.
There is a tight correlation between black hole mass and the stellar velocity dispersion
of bulges in the present-day universe (Ferrarese \& Merritt 2000; Gebhardt et al. 2000),
suggesting the growths of black holes and galaxies are
closely connected. The rapid decline of star formation rate and
quasar activity for the last $\sim$ 8 billion years also indicates the co-evolution of
black holes and galaxies (Dunlop 1999;
%Wyithe \& Loeb 2002;
Wolf et al 2003).
The feedback from Active Galactic Nuclei (AGNs) in star-forming galaxies
could quench further star formation and black hole growth at the same time
(Silk \& Rees 1998; Wyithe \& Loeb 2003; Springel et al. 2005).
To better understand the role of black holes in the AGN-galaxy connection,
we are carrying out detailed investigations of active supermassive black holes
% on the one hand,
and the evolution of their host galaxies.

The relation of AGN activity to black hole mass is important to investigate
since the black hole mass
sets the scale for the gravitational potential and also shows
%is one of the fundamental parameters in AGN energetics, which shows
the integral of the accretion history of the black hole.
A naive linear scaling between black hole mass and AGN luminosity,
expected from AGNs accreting at a fixed Eddington ratio,
is not observed (Woo \& Urry 2002a; O'Dowd et al. 2002).
%seems to be against the observation (Woo \& Urry 2002a).
However, black hole mass estimates with various indirect methods are somewhat uncertain
and more accurate data for various types of AGNs at various redshift ranges
are needed.

Studies with {\it Hubble Space Telescope (HST\footnote{
Based on observations made with the NASA/ESA {\it Hubble Space Telescope},
obtained from the data archive at the Space Telescope Institute.
STScI is operated by the association of Universities for Research
in Astronomy, Inc. under the NASA contract  NAS 5-26555.
})} data
showed that host galaxies of bright quasars
are typically massive ellipticals with a de Vaucouleurs profile
(Taylor et al. 1996; Urry et al. 2000; Dunlop et al. 2003),
%These early type host galaxies follow the Kormendy relation between the half-light
%radius ($r_{e}$) and the surface brightness ($\mu_{e}$),
structurally indistinguishable from normal galaxies.
%indicating structural similarities between normal and AGN host galaxies.
%The colors of host galaxies at low redshift also seem to be consistent with normal early type galaxies
%(Dunlop et al. 2003). At higher z, however, there are some indications of
%bluer colors than in normal galaxies at the same redshift (XXXX) while other studies shows
%no evidence of color difference (XXX).
Massive early-type galaxies hosting
BL Lac objects and radio galaxies
in the local universe ($z <0.1$)
seem to lie on the same fundamental plane as normal
galaxies (Falomo et al. 2002; Barth et al. 2003; Woo et al. 2004).
Furthermore, Woo et al. (2004) showed that the mass-to-light ratio evolution of AGN host galaxies
out to $z \sim 0.3$ is similar to that of normal galaxies,
showing that major star formation in the host galaxies occurs earlier than $z \sim 1$,
consistent with the mass-to-light ratio evolution of normal early-type galaxies.

To further investigate the relation of AGN activity to black hole mass, the properties of AGN host galaxies,
and the relation of active nucleus to host galaxy,
we undertook a program of absorption-line spectroscopy for a sample of AGN host galaxies.
From stellar velocity dispersions we can infer the black hole mass and study the fundamental relations
between black hole mass and other AGN properties.
With the addition of morphological information, we can also investigate the evolution
of AGN host galaxies using mass-to-light ratios. First results reported
by Woo et al. (2004), who focused on the fundamental plane of host galaxies.
Here, we present stellar velocity dispersions, black hole mass estimates, and
the mass-to-light ratio evolution for 39 AGN host galaxies
(32 BL Lac objects and 7 radio galaxies) out to $z \sim 0.6$,
including our previous work.
Using the structural parameters of the host galaxies and the
fundamental plane relation, we also estimate additional black hole masses for 34 BL Lac objects,
to investigate the role of black hole mass in the blazar unification paradigm.              

In \S~2, we describe the observations and velocity dispersion measurements, and
in \S~3, the AGN black hole mass and its relation with other AGN properties.
In \S~4, we discuss the evolution of AGN host galaxies, and
in \S~5 we present the discussion and conclusions.
We adopt a cosmology with $\Omega=0.3$, $\Lambda=0.7$, and $H_{0}=70$ km sec$^{-1}$ Mpc$^{-1}$.

\section{Observations and Data Reduction}

We selected $\sim 28$ BL Lac objects and radio galaxies at $z \lesssim 0.6$
with available $HST$ images to measure stellar velocity dispersions and mass-to-light ratios.
Detailed sample selection and data analysis for the first 15 host galaxies
can be found in Woo et al. (2004).
%We were able to measure the stellar velocity dispersions for 28 host galaxies.
Here, we briefly summarize the observations, data reduction, and velocity dispersion measurements for an additional 13 sources at relatively higher redshift.

\subsection{Observations and Data Reduction}

The spectra were obtained with the B\&C long-slit spectrograph at
the 6.5-m Magellan Clay Telescope at Las Campanas Observatory and
with the GMOS at the Gemini-South 8-m Telescope.
Table 1 shows the details of instrumental setups and the journal of observations.
The instrumental setups were chosen to cover strong stellar
absorption lines, such as G-band (4300 \AA), Mgb triplet (around 5172 \AA), and
Ca+Fe (around 5269 \AA),
and to provide sufficient instrumental resolution.
%$\lesssim 100$ km sec$^{-1}$ (Gaussian dispersion), so that the expected
%velocity dispersions ranging 150 to 350 km sec$^{-1}$
%for our target galaxies could be reliably recovered.
Sky conditions were mostly photometric.% with excellent seeing.
%Total exposure times of several hours, devided into 30 minutes exposures,
%were used depending on the host galaxy magnitude and the redshift.

The standard data reduction procedures, such as bias subtraction, flat-fielding,
spectral extraction, and wavelength calibration, were performed with IRAF routines.
One-dimensional spectra were extracted from each exposure
%using a 2.5$^{\prime\prime}$ aperture size.  Extracted spectra were
and combined to make the final spectrum for the velocity dispersion and redshift measurements.

\subsection{Dispersion Measurements}

We used a direct fitting method, in which the observed spectrum
is directly fitted in pixel space with broadened template spectra
(van der Marel 1994; Rix et al. 1995;
%Kelson et al. 2000a;
Barth et al. 2002; Woo et al. 2004).
The best-fitting dispersion value was determined by minimizing $\chi^{2}$
for the fit.
%For the fitting procedure featureless continua with various
%slopes were added to the template star spectra to account for the possible
%presence of an AGN non-stellar continuum, which changes the
%line strength as a function of wavelength.
The extracted galaxy and template star spectra were first normalized by a continuum fit.
%and re-binned logarithmically.

The template spectra, taken with each instrument, were convolved with Gaussian velocity profiles,
and fitted to the normalized galaxy spectrum using the
Gauss-Hermite Pixel Fitting software\footnote{available at http://www.stsci.edu/$\sim$marel/software.html}
(van der Marel 1994).
The fitting software uses various polynomial orders
and line strength parameters to match galaxy spectra,
and determines the best $\chi^{2}$ fit, which gives
the velocity dispersion measurement.
Galactic absorption lines and various AGN emission lines
(e.g., clearly present H${\beta}$ and [OIII] lines)
were masked out before fitting.

Extensive and careful fitting in various spectral regions
%, such as G-band, H${\beta}$, Mgb triplet, and NaD line regions,
was performed to determine the best-fitting spectral range.
%Various polynomial orders and continuum slopes were
%used for fitting to test any dependence on the parameters.
%Deviations with these parameters are similar to typical measurement errors.
Using template stars with different spectral types gives a larger variation in the velocity
dispersion. After fitting with each individual and combined template spectrum,
we chose the best-template star with the smallest $\chi^{2}$ for each galaxy.
Figure 1 shows the host galaxy spectra with the best-fitting broadened templates.

The measured velocity dispersions ($\sigma$) are corrected for differences in
instrumental resolution between the galaxy and template spectra
using Eq. 1 in Woo et al. (2004). The instrumental resolution correction changes
the velocity dispersion by a few percent.
The corrected velocity dispersions and aperture radii for 13 AGN host galaxies
are summarized in Table 2.

\subsection{New Redshift Measurements}

We measured the redshift for all observed host galaxies, which include 13 new objects
in addition to the 15 reported in Woo et al. (2004).
Radial velocity template stars were used to fit overall spectral features
including many absorption lines over a wide range of wavelengths.
Our high signal-to-noise ratio spectra give very accurate redshifts
with typical errors less than 0.0001.
Values for some redshifts given in the literature turned out to be wrong or inaccurate
probably due to lower data quality and sometimes misidentification.
In Table 3 we give all updated redshifts for our velocity-dispersion-measured AGNs.

We report new redshifts for two BL Lac objects.
% 1248-296.
% and  1133+161, and for the radio galaxy 3C306.1.
An uncertain redshift of 0.487 was previously reported for 1248-296
(Padovani \& Giommi 1995);
its true redshift is $0.3819 \pm 0.0001$ from fitting many absorption lines
around the G-band.
For 1133+161, $z= 0.46$ was reported by Fichtel et al. (1994)
%while Rector et al. (2000, AJ, 120, 1626)
while Rector et al. (2000) estimated a much larger redshift
although they reported it as tentative since absorption lines could not be clearly
identified in their spectrum.
We report $z = 0.5735 \pm 0.0001$ for 1133+161.
%The redshift of radio galaxy 3C306.1 was measured with narrow emission lines
%to be $z = 0.441$ (Spinrad et al. 1985).
%However, AGN emission lines seem to be redshifted compared to host galaxy absorption lines.
%We report $z=0.4194 \pm 0.0001$ based on stellar absorption lines around G-band,
%and average $z=0.4402$ from AGN emission lines including Balmer ($H\beta$, $H\gamma$, $H\delta$) and
%[OIII] lines.

%We actually observed more BL Lac objects, but failed to measure the stellar velocity dispersion
%because of too bright continuum. In these cases, however, we were easily able to measure the redshift.
%XXXXX

\section{AGN Black Holes}
\subsection{Black Hole Mass Estimation}

%Black holes mass is one of the most fundamental paremeters in AGN physics.
The reverberation mapping technique gives reliable black hole mass estimates (Peterson 1993),
but this method is very expensive, requiring long-term monitoring. An indirect
method, using the scaling of the size of the broad-line region with UV/optical luminosity
(Kaspi et al. 2000)
is also popularly used for black hole mass estimation.
However, this method has a large scatter, limited luminosity range, and in any case
can be applied only for broad-line AGNs.

The correlation between black hole mass and galaxy luminosity or stellar velocity
dispersion opens a new way of estimating black hole mass for AGNs, including
Type II AGNs, radio galaxies, and BL Lac objects.
Since the host galaxy magnitude is relatively easy to measure using images with a high quality point spread function,
the black hole mass-galaxy luminosity relation can be a powerful tool for estimating black hole mass
for a large sample of AGNs. However, scatter in the mass-to-light ratio among galaxies
increases the uncertainty. Also, because the black hole mass-galaxy luminosity relation is derived from a
local galaxy sample, a correction for luminosity evolution of the host galaxies is necessary (Woo et al. 2004),
which contributes additional uncertainty.

The black hole mass-stellar velocity dispersion relation is much tighter than the
mass-luminosity relation in local galaxies,
and it plausibly holds up to
%a galaxy formation epoch.
an epoch when the bulk of the galaxy mass has assembled, well above the redshift discussed here.
Our sample consists of much higher redshift AGNs (out to $z \sim 0.6$)
than previous samples studied with the mass-dispersion relation (Barth et al. 2002; Falomo et al. 2003; Woo et al. 2004). However, they are
still relatively low-redshift AGNs with very massive host galaxies ($> 10^{11} M_{\odot}$) so it is reasonable
to use the mass-dispersion relation; certainly,
the mass-dispersion relation for different AGN types and redshift ranges, especially at high redshift ($z \gtrsim 1$) should be investigated further (cf. Treu et al. 2004; Silge et al. 2005).

We use the measured stellar velocity dispersion of host galaxies to estimate
the black hole mass via the mass-dispersion relation of Tremaine et al. (2002):
\begin{equation}
M_{\bullet} = 1.349 \times 10^{8} M_{\odot} (\sigma_{e}/200~ {\rm km~s}^{-1})^{4.02} ~,
\label{sig}
\end{equation}
with proper aperture corrections for
%luminosity weighted velocity dispersion within $r_{e}$ on the basis of averaged velocity  
$\sigma_{e}$ following Jorgensen et al. (1995).
Table 3 presents the black hole masses for our sample of 39 AGNs (which consists of 32 BL Lac objects and 7 radio galaxies; 28 of these we observed ourselves; the remaining 11 have published velocity dispersions from other authors, as summarized in Table~3 of Woo et al.~(2004)).

In Figure 2, we compare black hole mass estimates from the measured $\sigma_{e}$
and other, likely less reliable, mass estimates from the calculated $\sigma_{e}$
using $r_{e}$ and $\mu_{e}$ via the Coma cluster fundamental plane.
The rms scatter between two black hole mass estimates is 0.34 dex
after correcting the luminosity evolution of stellar populations
%The surface brightness ($log~I$) is corrected for the averaged luminosity evolution of stellar populations
with $\Delta log~L/\Delta log~z =0.502$ in the rest-frame Cousins R-band from our host galaxy evolution study (see \S 4.3).
Without this evolutionary correction, the black hole masses are
%more uncertain ( the rms scatter $\sim 0.4$ dex) and
systematically higher
because of the correlation between the $log~\sigma$ and $log~I_{e}$ in the fundamental plane relation.
%the scatter is larger (0.45 dex) and
Although the intrinsic scatter of the fundamental plane
and the scatter in the luminosity evolution
will increase the uncertainties
%, in addition to the scatter
above those in the mass-dispersion relation,
estimating black hole mass from the $r_{e}$ and $\mu_{e}$ could be
an alternative way for AGNs with particularly luminous nuclei,
where host galaxy stellar velocity dispersions are difficult to measure.
In that case, the $K$-correction and the luminosity evolution correction of host
galaxies need to be done carefully.

\subsection{The Black Hole Masses and Eddington Ratios of BL Lac Objects}

BL Lac objects are AGNs with a relativistic jet oriented toward
the line of sight (Urry \& Padovani 1995).
%Because of the beaming effect, they can be observed at much lager redshift.
Depending on the wavelength of the two broad peaks in their spectral energy distributions
(SEDs),
they are classified as low-frequency-peaked BL Lac (LBL) or high-frequency-peaked BL Lac (HBL) objects; in particular the classification can be done
based on the X-ray to radio flux ratio.\footnote{We use the definition of high (low)-frequency-peaked BL Lac with a dividing line
at $log f_{X}/f_{r} = -5.5$ (
%Wurtz 1994;
Perlman et al. 1996),
with X-ray flux at 1 KeV and radio flux at 5~GHz in Janskys.}
The physical cause of the different SED shapes of BL Lac objects has been the subject of many studies.
LBLs from radio surveys are typically more luminous than HBLs from X-ray surveys.
This is interpreted as either that LBLs are more beamed (an orientation effect: Ghisellini \& Maraschi 1989;
Urry \& Padovani 1995) or that LBLs are intrinsically more luminous
(Padovani \& Giommi 1995; Fossati et al. 1997; 1998; Ghisellini et al. 1998).

Here we study the AGN engine for a sample of 32 BL Lac objects with black hole masses
from measured stellar velocity dispersions (Table 3).
In addition, we collected 34 BL Lac objects as a supplementary sample
with known redshifts and host galaxy magnitudes (Urry et al. 2000),
for which we estimate black hole masses from $r_{e}$ and $\mu_{e}$ via the fundamental plane.
X-ray and radio fluxes are collected from the literature using the NED database.\footnote{
The NASA/IPAC Extragalactic Database (NED) is operated by the Jet
Propulsion Laboratory, California Institute of Technology,
under contract with the National Aeronautics and Space Administration.}

Figure 3 shows black hole mass estimates and the X-ray to radio flux ratio
for these samples.
The black hole mass of the dispersion measured sample ranges from $\sim 4 \times 10^{7}$ to
$\sim 6 \times 10^{8} M_{\odot}$,
similar on average to that of radio galaxies and bright quasars,
but with a much narrower range (Woo \& Urry 2002a).
The lack of higher or lower black hole
masses seems to be the result of selection effects.
The volume for the dispersion measured sample is too small ($<z> \sim 0.17$ with a standard deviation $0.12$)
to contain the more massive black holes at higher redshifts
although we cannot rule out that the black hole mass upper limit of BL Lac objects
is much lower than that of other AGNs.
When we include the less reliable black hole mass estimates of the supplementary sample,
which occupies a larger volume ($<z> \sim 0.31$ with a standard deviation $0.17$),
the largest black hole mass increases to $\sim 4 \times 10^{9} M_{\odot}$.
Another selection effect is that less massive black holes ($\lesssim 10^{7} M_{\odot}$)
tend to have fainter host galaxies,
which could be below the detection limit of the HST imaging snapshot survey.

The black hole mass ranges for LBLs and HBLs are similar,
as was found previously with much smaller samples
(Falomo et al. 2002; Barth et al. 2003; Woo et al. 2004), indicating that black hole mass is not
the physical parameter determining BL Lac SED types.
Black hole mass also does not
%strongly
correlate with either X-ray or radio luminosity, as shown in Figure 4.
LBLs ({\it circles}) tends to have higher radio luminosities,
almost by definition, but with similar black hole masses as HBLs ({\it triangles}).
The mean black hole masses of two types are
%log M$_{LBL} = 8.29 \pm 0.60$ and log M$_{HBL} = 8.30 \pm 0.40$.
log M$_{LBL} = 8.29 \pm 0.13$ and log M$_{HBL} = 8.30 \pm 0.06$.

We also calculated bolometric luminosities (without beaming correction)
using the radio to bolometric luminosity relation, derived from the blazar sample of
Fossati et al.~(1998), for which bolometric luminosities were integrated from the SED models as a function of radio luminosity.
Figure 5 shows the relation among black hole mass, bolometric luminosity, and Eddington ratio.
For a given black hole mass, there are 2--3 orders of magnitude difference in Eddington ratio, with
LBLs generally showing a higher Eddington ratio.
This can be interpreted as LBLs have
higher apparent Eddington ratio in radio and in bolometric luminosity than HBLs,
either because of more beaming or higher intrinsic power (Urry \& Padovani 1995).
We cannot differentiate between these two scenarios without an accurate beaming correction which
is not possible for individual objects with precision better than an order of magnitude.

\subsection{Radio Power of Blazars}

Since the correlation of black hole mass with radio power was first suggested
for a handful of galaxies (Franceschini et al. 1998), several
studies have attempted to demonstrate such a correlation (McLure et al. 1999; Lacy et al. 2001; Jarvis \& McLure 2002).
However, most of the studied samples seem to suffer from selection effects.
Woo \& Urry (2002a, 2000b) showed that the black
hole mass ranges are not different between radio-loud and radio-quiet samples with
over 400 AGNs.
It has since been shown for a much larger sample of Sloan Digital Sky Survey AGNs ($\sim 6000$)
that for a large black hole mass range ($10^{7} \lesssim M_{\bullet} \lesssim 10^{10}$),
the radio-loudness parameter ($F_{5GHz}/F_{B}$) spans more than 4 orders of magnitude
(McLure \& Jarvis 2004, see Figure 2 in their paper),
although the mean black hole mass of radio-loud AGNs is a factor of $\sim$ 1.6 larger
than that of radio-quiet AGNs.

If radio-loud and radio-quiet AGNs are very different populations with    
different central engines, then a correlation between black hole mass and radio power
might exist only among radio-loud AGNs.
We compare black hole mass with radio power for our sample of BL Lac objects
and for flat-spectrum radio quasars (FSRQs) from Oshlack et al. (2002).
The black hole masses of the FSRQs were estimated from the broad-line width and optical luminosity
(Oshlack et al. 2002; Woo \& Urry 2000a).
If the distribution of BLR clouds is not random and more like a disk distribution,
then the velocity of the BLR clouds could be higher by factors of a few (Jarvis \& McLure 2002),
and thus the black hole masses larger by as much as an order of magnitude.

Figure 6 shows the black hole mass and radio luminosity at 5~GHz for BL Lac objects
({\it circles and triangles}) and FSRQ ({\it crosses}).
The radio luminosity is not beaming corrected
and the intrinsic radio powers of these AGNs are much lower.
%Assuming the highest value of Rolentz factor (10), intrinsic radio power will
%3 orders of magnitude low.
However, it is clear that radio luminosity between
FSRQs and BL Lac objects is different by a minimum of several orders of magnitude
for the given black hole mass range.
%of the BL Lac object sample ($7.5 \lesssim log~M_{\bullet} \lesssim 9$).
It is unlikely that any beaming correction would reveal
a hidden correlation between black hole mass and radio luminosity. Even after increasing
the black hole mass of FSRQs by an order of magnitude,
considering the possibility of underestimation because of the uncertainties in BLR
cloud distribution (Jarvis \& McLure 2002), the radio luminosity
still spans over 4 orders of magnitude, indicating no strong correlation between black hole mass
and radio power.
That is, BL Lac objects and FSRQs may differ in radio luminosity but not in black hole mass,
indicating BL Lac objects have lower Eddington ratios than FSRQs.
This is consistent with the view that FSRQs and BL Lac objects are the same fundamental
class of AGN with different intrinsic radio and line luminosities (Padovani 1992; Maraschi \& Tavecchio 2003).
One FSRQ, PKS 0921-213, has relatively low radio power compared with other FSRQ.
This quasar was identified as a double-peaked emission line source,
which probably has a considerably lower accretion rate (Eracleous \& Halpern 2003).

\subsection{Mass - Luminosity Relation for Radio-Loud AGNs}

The Eddington ratio represents how energetic a black hole is for its given mass.
Woo \& Urry (2002a) showed that the Eddington ratio spans up to 3 orders of magnitude
for given black hole mass of $\sim$ 300 AGNs.
Here, we revisit the mass-luminosity correlation of black holes for our sample of
BL Lac objects including radio-loud AGNs from Woo \& Urry (2002a).

It is very difficult to determine the Doppler factor for BL Lac jets
since superluminal motion and the beaming angle have to be measured for individual objects,
from VLBI imaging and constraints from multi-wavelength observations.
Good estimates of the Doppler factor do not exist for most of our objects.
We instead used an average beaming factor 3.9 for BL Lac objects from Dondi \& Ghisellini (1995),
who calculated a lower limit on the Doppler beaming factor for the $\gamma$-ray emission region.

Figure 7 shows bolometric luminosity and black hole mass for radio-loud AGNs
including 66 BL Lac objects ({\it circles}).
%Because of the small volume we covered for the BL Lac objects
%($<z> \sim 0.17 \pm 0.12$), the black hole mass range is much narrower compared with that of radio-loud AGN.
It clearly shows that BL Lac objects are intrinsically low-power AGNs with
lower Eddington ratios.
The Eddington ratio spans at least 4 orders of magnitude, given that the
Doppler-factor-corrected luminosity of each BL Lac object is an upper limit.
AGN luminosity continuously goes down to non-active levels, however, 
fainter non-beamed quasars and brighter BL Lac objects 
do not apper in this plot because of the flux limit and the limited volume, respectively, resulting in a spurious bimodal distribution of bolometric luminosities.

\section{Host Galaxy Evolution}

We studied the evolution of host galaxies
for a total sample of 39 radio-loud AGNs: 7 radio galaxies and 32 BL Lac objects.
We first measured basic galaxy properties from spectroscopic and imaging data,
then computed the mass-to-light ratio for each galaxy and investigated
the evolution of the mass-to-light ratio out to $z \sim0.6$.

\subsection{Host Galaxy Properties}

High signal-to-noise ratio spectra for all our sample galaxies are available (Falomo et al. 2002;
Barth et al. 2003; Woo et al. 2004; this work)
and the central stellar velocity dispersions are measured with an aperture corresponding to
$3.4^{\prime\prime}$ at the distance of the Coma cluster.
The half-light radius, $r_{e}$, and the total galaxy magnitude in the Cousins $R$ band are mainly taken from
the HST snapshot survey of BL Lac objects (Urry et al. 2000; see Woo et al. 2004 for other sources).
The mean surface brightness within $r_{e}$ is derived from the
total magnitude of the host galaxy using:

\begin{equation}
<\mu_{e}> = (m -K) + 5 log(r_e) + 2.5 log(2\pi) - A_R -2.5 log (1+z)^{4} ~,
\end{equation}
where $m$ is the observed galaxy magnitude,
$K$ is the $K$-correction value,
$r_{e}$ is the effective radius in arcseconds,
and $A_R$ is the foreground extinction in the $R$ band taken from the NED database.
%%(Schlegel et al. 1998).
Initially, we derived the observed-frame $<\mu_{e}>$ without $K$-correction (\S~\ref{obsML}) and
after determining the best-fitting star formation redshift, we calculated
the rest-frame $<\mu_{e}>$ with $K$-correction values
from our SED models (\S~\ref{intML}). In Table 2, we list galaxy properties
for the 13 additional sources with new spectroscopic observations presented
in this paper.

There are 4 radio galaxies for which $r_{e}$ and total galaxy magnitude measurements from HST data are not available.
%For 3C 348, we got sersic index=1.33 and $r_e=4.8''$ for n=4 we got $r_e=39.0''$
For 3C 348, we adopted Harris V=16.36 and $r_e=22.96''$
based on ground-based telescope data (Roche \& Eales 2000).
Because
%Assuming
the Harris $V$ and the Johnson $V$ have very similar response functions,
we convert the $V$ into the Cousins $R$ using $V-Rc=0.723$,
interpolated for its redshift $z=0.154$, from Fukugita et al. (1995), which is similar
to $V-Rc=0.721$ from our population synthesis models with $z_{form}=2$.
%to V-Rc=0.713 from our population synthesis models with $z_{form}=1.6$.

For the other three radio galaxies, namely 3C 135, 3C 424, and 3C 306.1,
we could not find any host galaxy magnitude and $r_e$ measurements
from the literature and thus measured them
from $HST$ images using GALFIT (Peng et al. 2002).
PSF images, generated using the Tiny Tim software (Krist \& Hook 1997),
were convolved with $HST$ images and the best fit Sersic index and radius
were measured. We found that the derived Sersic indices were close
to de Vaucouleurs profiles.
Hence, we fixed the Sersic index to 4 and derived $r_{e}$ and the total galaxy magnitude. Galaxy properties for the 13 new AGN host galaxies are presented in Table 2.

\subsection{Observed Mass-to-Light Ratio Evolution}
\label{obsML}

From the virial theorem, galaxy masses can be estimated as
\begin{equation}
log~M = 2 log~\sigma + log~r_{e} + C_{1},
\label{hostM}
\end{equation}
where $\sigma$ is in km s$^{-1}$, $r_{e}$ in $Kpc$, M in $M_{\odot}$ units, and $C_{1}$ is 6.07 (Bender et al. 1992).
Log indicates the base 10 logarithm.
The average surface brightness within $r_{e}$ is defined by
\begin{equation}
log~I_{e} \equiv  -0.4 (<\mu_{e}> +C_{o}),
\end{equation}
in units of $L_{\odot}~pc^{-2}$,
where $\mu_{e}$ is the average surface brightness within $r_{e}$
in $mag~ arcsec^{-2}$ and $C_{o} = -26.40$ for the Gunn $r$ band
(Jorgensen et al. 1996) and -26.05 for the Cousins R band, taking $R
%= 4.42$ for the Sun (Binney \& Merrifield 1998).
= 4.48$ for the Sun (Worthey 1994).                  
Since the luminosity of a galaxy is defined by $L = 2 \pi r_{e}^{2} I_{e}$,
the mass-to-light ratio can be expressed as
\begin{equation}
log~ M/L = 2~ log~\sigma - log~ I_{e} - log~ r_{e} +C_{2},
\end{equation}
where the constant $C_{2}= -0.73$ (Jorgensen et al. 1996).
% = log (K/ 2 \pi G)$.

The evolution of $log M/L$ is the difference in $log~I_{e}$
between two redshift points,
assuming the mass, $\sigma$, and $r_{e}$ remain the same.
At $z \sim 0$,  $log~I_{e}$ can be derived from the
fundamental plane of early type galaxies:

\begin{equation}
log~ I_{e} = (b~log~ r_{e} + c~log~ \sigma + \gamma)/a.
\end{equation}
Here, a=0.82, b=-1, c=1.24, $\gamma=0.182$, as derived for the Coma cluster fundamental
plane in the Gunn $r$ band with $r_{e}$ in arcseconds (Jorgensen et al. 1996).
If instead $r_{e}$ is expressed in kpc, then $\gamma = -0.120$ if one assumes a Coma distance of 102.9 Mpc.
After transforming the fundamental plane of the Coma cluster to the Cousins $R$ band, the evolution of the M/L ratio in the $R$ band can be written as follows
\begin{equation}
\Delta log~ M/L =  (b~log r_{e} + c~log \sigma + \gamma) /a +0.4(<\mu_{e}(z)> + (r-R) +C_{o}).
\end{equation}
Here, the Cousins $R$ band surface brightness, $<\mu_{e}(z)>$, is from Eq. 2,
$C_{o} = -26.40$ for the Gunn $r$ band, and the color of elliptical galaxies in Coma is $r-R \approx 0.35$ (Jorgensen 1994).

The mass-to-light ratio is usually derived for a rest-frame broad-band magnitude.
For high redshift galaxies, it is necessary to apply a $K$-correction
to get the rest-frame magnitude from the observed magnitude.
The $K$-correction is typically derived
from template or model spectra assuming a cosmological model (Fukugita et al. 1995; Poggianti 1997).
These models, however, already include an assumption of the galaxy formation epoch.
Thus, the $K$-corrected mass-to-light ratio suffers uncertainties in  constraining galaxy formation and evolution.
We therefore decided to use the mass-to-light ratio in the observed frame without $K$-correction to constrain the star formation epoch (i.e., we use equations2 and 7, but with $K$ set to zero).
The advantage of using observed-frame M/L ratio evolution is that an a priori assumption on
the star formation epoch can be avoided and model predictions with different
formation epoch ($z_{form}$) show larger differences in the M/L ratio evolution,
especially at low redshift.

The evolution of the observed M/L ratio for our sample of AGN host galaxies is measured individually and
averaged at each redshift bin. We redefine $\Delta log~ M/L$ as the difference in logarithm
between the M/L ratio at a certain redshift and the M/L ratio in our lowest redshift bin at $z=0.046$ (Figure 8). So we compare AGNs at different redshift to each other instead of comparing directly to the normal galaxies in Coma. This relative comparison has the advantage that it avoids potential systematic errors resulting from differences in absolute calibration between our AGN dataset and the Coma dataset of Jorgensen et al.~(1996). It also removes from the comparison any dependency on the actual distance to Coma.
Figure 8 shows that as z increases, the observed-frame M/L remains approximately
constant. The stellar populations do get younger and intrinsically
brighter. However, this is approximately cancelled by two other
effects. First, as z increases, the observed-frame band corresponds
to a bluer rest-frame band. Since relatively old stellar populations
are redder than the Sun, they have lower $L/L_{\odot}$ in bluer bands. And
second, due to the (1+z) stretching of the spectrum, the observed
frame samples a smaller range in wavelength in the rest-frame as
z increases.
%As z increases, stellar populations get younger and intrinsically
%brighter, but because of the redshift effect (mostly from the 1+z stretch), the observed
%magnitudes are roughly flat (Figure 8).

We constructed SED models with single burst star formation epochs ($z=1-5$)
using Bruzual \& Charlot (2003) models with the Salpeter IMF and the solar metallicity.
We then redshifted the models to produce the observed magnitude at each redshift.

Figure 8 shows that the averaged M/L ratio values over each redshift bin
({\it filled circles}) is consistent with passive evolution models
with $z_{form}=1-2$. We found that single burst models with $z_{form}=1.4^{+0.9}_{-0.2}$ best
reproduce the observed M/L ratio evolution
within the 68\% confidence limit based on $\chi^{2}$ analysis.
The last redshift bin has only two points and one of them, 3C 306.1
has clear dust lanes. If we exclude the last bin,
then the best fit becomes $z_{form}=1.4^{+0.7}_{-0.2}$. Therefore, we simply used the $z_{form}=1.4$ model for
further analysis.
The derived star formation epoch for our sample of AGN host galaxies is
consistent with that of normal
galaxies in the field and clusters ($z_{form} = 1$--3; van Dokkum \& Franx 2001; Treu 2002; Rusin et al. 2003; van Dokkum \& Stanford 2003),
indicating normal and AGN host galaxies experience similar formation histories.

\subsection{Intrinsic Mass-to-Light Ratio Evolution}
\label{intML}

In order to derive the evolution of the mass-to-light ratio in the rest-frame $R$ band,
we calculated $K$-correction values from our passive evolution model
with the determined star formation epoch, $z_{form}=1.4$.
Figure 9 shows the difference in $K$-correction value in our models
with different $z_{form}$ and popularly used models from Fukugita (1995)
and Poggianti (1997).
The $K$-correction is the difference between the observed magnitude
from the redshifted spectrum and the rest-frame magnitude from the de-redshifted spectrum.
The $K$-correction value decreases as the formation redshift decreases
%because the stellar population age difference is decreasing
%between $z=0$ and a given redshift.
because stellar populations get younger at a given redshift
and thus the observed magnitude decreases.
Note that popularly used models
predict large $K$-corrections owing to their older adopted cosmology.
The $K$-correction can significantly change the intrinsic mass-to-light ratio estimation.
At $z=0.5$, the difference in $K$-correction between our $z_{form}=1.4$ model
and the popularly used models corresponds to $\sim 0.05$ in $log~M/L$.
We provide $K$-correction values for a couple of single burst stellar population models in Table~4--7.

With our best-fit $K$-correction value, we derived rest-frame $R$ band
magnitudes and the mass-to-light ratios for individual galaxies.
Figure 10 shows the mass-to-light ratio evolution in the rest-frame Cousins $R$ band for our AGN host galaxies.
The averaged mass-to-light ratio indicates 50\% increase in M/L ratio between $z=0.05$ and $z=0.4$.
%,
which corresponds to $\Delta log (M/L)/ \Delta z = -0.502$ in the Cousins $R$ band and
$\Delta log (M/L)/ \Delta z = -0.619$ in the B band based on our population synthesis model. The lowest and highest $z_{form}$ models consistent with the data at 68\% confidence, as determined in the previous
section ($z_{form}=1.2$ and 2.3), indicate a 68\% confidence range for the
M/L evolution between -0.581 and -0.421 in the Cousins $R$ band, and between -0.695
and -0.517 in the $B$ band. The trend of the M/L evolution of our AGN host
galaxies is also similar to that of normal galaxies with  $\Delta log (M/L)/ \Delta z = 0.46-0.72$ in the $B$ band (Treu et al. 2002; Rusin et al 2003; van de Ven et al. 2003; van Dokkum \& Stanford 2003).

\subsection{Mass vs. M/L relation}

We derived the correlation between galaxy mass and mass-to-light ratio
(after correcting for luminosity evolution
with  $\Delta log (M/L)/ \Delta z = -0.502$) for the AGN host galaxies.
The results can be fit with
\begin{equation}
log~ M/L = (0.45 \pm 0.05) \times log~M -4.39(\pm 0.57) .
\label{eqMML}
\end{equation}
The AGN host galaxies are shown as filled squares in Figure 11, and are compared with
nearby early-type galaxies ({\it open circles}) from
van der Marel (1991), Magorrian et al. (1998), Kronawitter et al. (2000),
and Gebhardt et al. (2003).
The mass-to-light ratio of the nearby galaxies
are all based on detailed dynamical models for spatially resolved
kinematical data (this is more detailed than the
values for our own sample, which are based on the virial theorem and
an assumption of homology). The compilation into a homogeneous set
is from van der Marel \& van Dokkum (2005). All mass-to-light ratio values were
transformed to the Cousins $R$ band using $B-R=1.57$ (Fukugita et al. 1995).
Distances for the large
majority of the nearby galaxies were taken from Tonry (2001) with $H_{0} = 70$ km sec$^{-1}$ Mpc$^{-1}$.
Data for the same galaxies from different authors were
averaged. Only galaxies classified as ellipticals are included in the nearby galaxy sample. The
few most distant galaxies from Magorrian et al. (1998) were removed
from the sample because their models for these galaxies included
unrealistically large black hole masses.
We find that normal and AGN host galaxies have a similar relation between mass and
M/L ratio in the galaxy mass range where the data sets overlap, $11 < log~M/M_{\odot} < 12$. The most massive galaxies in the AGN sample suggest the possibility of a break in the $M/L$ vs.~$M$ relationship at masses in excess of $10^{12} M_{\odot}$.

\section{Discussion and Conclusions}

We measured the stellar velocity dispersions for 21 BL Lac object host galaxies and 7 radio galaxies
from our spectroscopic observations.
Including 11 velocity-dispersion-measured BL Lac object host galaxies from the literature,
we estimated black hole masses for a sample of 39 AGN host galaxies using the mass-dispersion
relation.
We also estimated black hole masses for 34 additional BL Lac objects
using the derived velocity dispersion from well-measured $r_{e}$ and $\mu_{e}$ of the host galaxies
and the fundamental plane relation.

Estimating black hole mass from $r_{e}$ and $\mu_{e}$ seems promising since the intrinsic scatter in
the fundamental plane is small. Although high-resolution imaging is required for an accurate AGN subtraction
to derive reliable host galaxy properties,
it would be more feasible than long-exposure spectroscopy in the case of typical
quasars, of which a featureless AGN continuum is much brighter than host galaxy absorption features.
%Furthermore, black hole mass for
%bright quasars, for which measuring velocity dispersion from stellar absorption lines is extremely difficult,
%can be relatively easily estimated.

The black hole mass of BL Lac objects ranges from $ 10^{7}$ to $ 4 \times 10^{9} M_{\odot}$.
We found no strong correlation between black hole mass and either X-ray or radio luminosity.
HBLs and LBLs have similar black hole masses but LBLs show higher Eddington ratios in radio and bolometric
luminosity, because of either more beaming or higher intrinsic power.
We also compared FSRQs and BL Lac objects and found that their black hole masses are similar
but their radio luminosities are quite different, indicating that BL Lac objects and FSRQs
are plausibly the same objects with different Eddington ratios as suggested by
other blazar unification study (Maraschi \& Tavecchio 2003).

All black hole mass estimates depend on the mass-dispersion relation observed in the present-day universe.
A recent study on 7 Seyfert 1 galaxies at $z \sim 0.4$ suggests
an evolution of the mass-dispersion relation with a higher black hole mass
for a given velocity dispersion (Treu et al. 2004).
Neglecting the black hole mass growth for the last 4 billion years,
this could indicate a mass evolution of spheroids.
If this is the case, then our black hole mass estimates are a lower limit.
However, the host galaxies in our sample are very massive elliptical galaxies,
consistent with pure luminosity evolution since $z \sim 1$, and probably represent a different population than Seyfert 1 galaxies.

We measured the mass, mass-to-light ratio, and the evolution of the mass-to-light ratio for the sample
of 39 AGN host galaxies. From the observed-frame (no $K$-correction) mass-to-light ratio evolution,
we tested single burst star formation epoch models using our population synthesis models.
The passive evolution model with $z_{form}=1.4^{+0.9}_{-0.2}$ is consistent with the observed
mass-to-light ratio evolution.
%With $K$-correction from a passive evolution model with $z_{form}=1.4$,
From a passive evolution model with $z_{form}=1.4$,
we measured the evolution of the intrinsic mass-to-light ratio in the Cousins $R$ band,
which is $\Delta log (M/L)/ \Delta z = -0.502 \pm 0.08$. The mass-to-light ratio evolution of our sample of AGN host
galaxies is similar to that of normal galaxies, indicating that normal and AGN host galaxies
experience similar star formation histories. Whether the supermassive black hole is active
at the observed epoch seems not related to the global star formation history.
However, we note that our host galaxies are among the most massive galaxies
($> 10^{11} M_{\odot}$) and the star formation redshift is marginally lower
than that of normal galaxies with the same mass range ($z_{form} \gtrsim 2$--3;
Treu et al. 2005; van der Wel et al. 2005) indicating 1--2 Gyr younger age, which implies
either a later epoch of star formation or additional star formation
in AGN host galaxies. Rest-frame colors of host galaxies can shed light on
more detailed interpretation.

In contrast to host galaxies at high redshift ($z \gtrsim 2$),
when galaxies and black holes are still assembling their masses,
early-type host galaxies at low redshift ($z < 1$) are grown-up galaxies
with a typical mass $ \gtrsim 10^{11} M_{\odot}$.
These host galaxies seem just like normal galaxies except for their active central black holes,
which are probably revived from dormant status.
In the case of late-type host galaxies, where star formation and AGN activity can be more closely
connected, host galaxies might show very different properties compared to normal galaxies.
Further investigation of the relation of nuclear activity to host galaxy properties for AGNs
at higher redshift and with lower host galaxy mass is required to understand the full picture.

\acknowledgements
This research is a part of the AGN Key Project of the Yale-Calan collaboration, and has been
supported by Fundacion Andes.
Partial support was also provided by
NASA grant AST-0407295.
%We thank the referee for suggesting many constructive points.
%J. W. thanks the Departamento de Astronomia, Universidad de Chile
%for the hospitallity during the Yale-Calan collaboration meetings in Santiago.
Partly based on observations obtained at the Gemini Observatory,
which is operated by the Association of Universities for Research in Astronomy, Inc.,
under a cooperative agreement with the NSF on behalf of the Gemini partnership:
the National Science Foundation (United States),
the Particle Physics and Astronomy Research Council (United Kingdom), the
National Research Council (Canada), CONICYT (Chile), the Australian Research Council
(Australia), CNPq (Brazil) and CONICET (Argentina).

\clearpage
\begin{figure}
\plotone{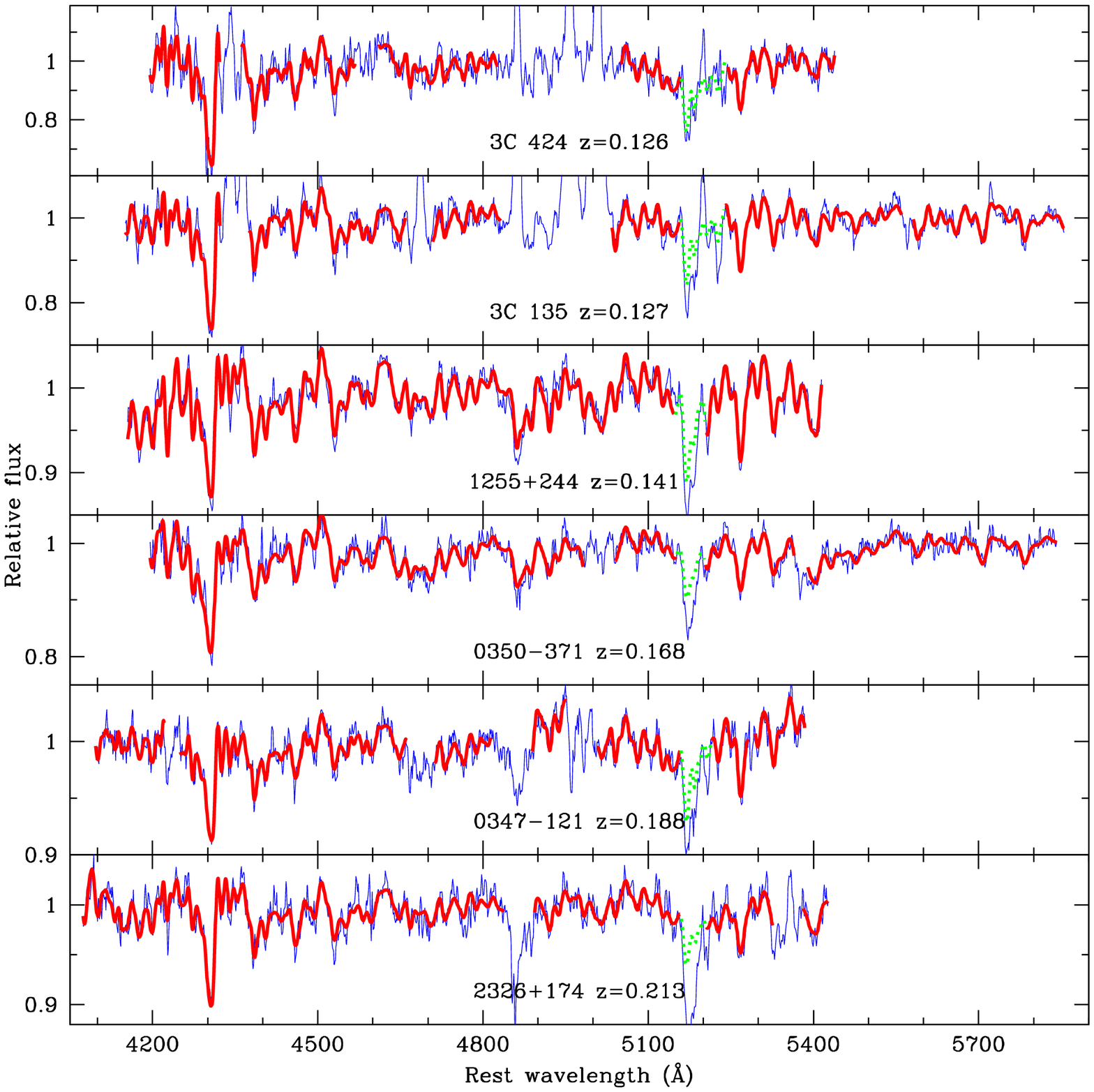}
\end{figure}

\begin{figure}
\plotone{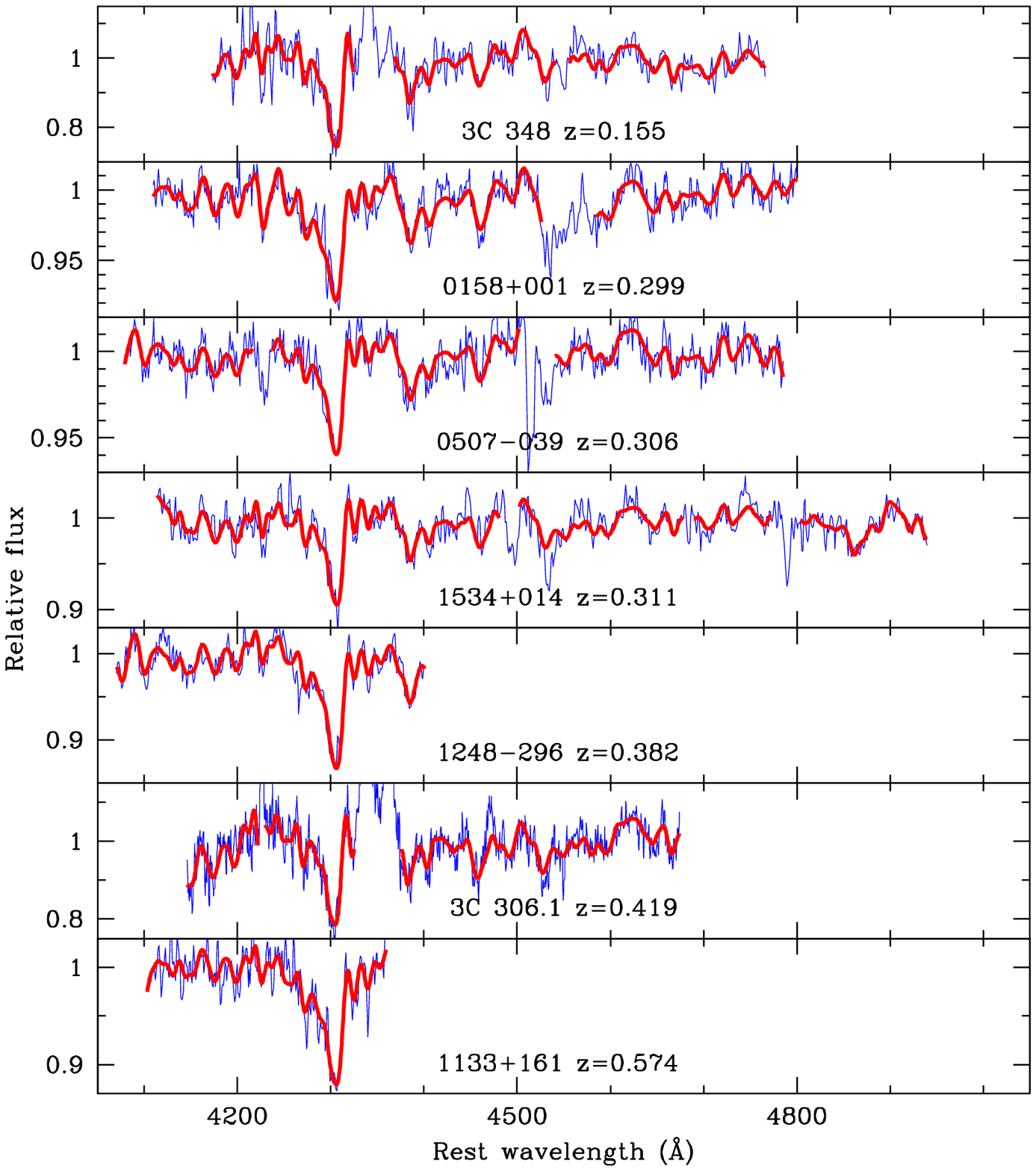}
\caption{
Observed spectra of AGN host galaxies
({\it thin line} [colored blue in electronic edition]),
with best-fit templates
({\it thick line} [colored red in electronic edition]).
{\it Top panels}: Spectra of six galaxies with large wavelength coverage.          
%Galaxy spectrum and the best-fit template are over-plotted for each galaxy.
Bad pixels, AGN emission lines, and the Mgb triplet lines
({\it dotted line} [colored green in electronic edition]) were masked
out before fitting.
{\it Bottom panel:} Seven galaxies with smaller fitting ranges.
}
\end{figure}

\begin{figure}
\plotone{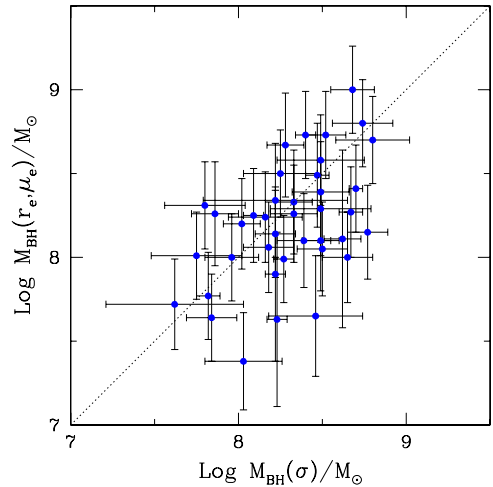}
\caption{Comparison of black hole mass estimates of 32 BL Lac objects
from the stellar velocity dispersions (Table~3) and from
$r_{e}$ and $\mu_{e}$ using the Coma cluster fundamental plane relation,
%$log~\sigma = (log~r_{e} +0.82~log~I_{e}+0.119)/1.24$
converted for a cosmology with $H_{o}=70~ km s^{-1} Mpc^{-1}$, $\Omega=0.3$, $\Lambda=0.7$. The surface brightness $log~I$ is corrected for the
rest-frame Cousins $R$-band luminosity evolution
of stellar populations using $dlog~L/dlog~z=0.502$ (see \S 4.3).
The measurement errors of $r_{e}$, $\mu_{e}$, and $\sigma$ are considered
in the error propagation. The intrinsic scatter in the fundamental plane (0.08 in log $r_{e}$)
is also included in the error estimation of black hole mass from $r_{e}$ and $\mu_{e}$ (Woo et al. 2004).
The rms scatter between two black hole mass estimates is 0.34 dex.}
\end{figure}

\begin{figure}
\plotone{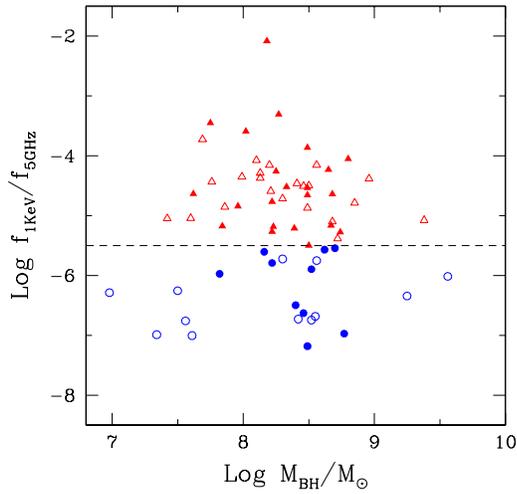}
\caption{X-ray to radio flux ratio vs. black hole mass for our sample of
BL Lac objects.
Black hole masses are estimated from the measured $\sigma_{e}$ (filled symbols) or
$r_{e}$ and $\mu_{e}$ (open symbols). The dashed line divides high-frequency peaked
BL Lac objects (triangles) [colored red in electronic edition] from low-frequency peaked objects (circles) [colored blue in electronic edition] following Perlman et al. (1996).
The black hole mass spans over 2 orders of magnitude independent of BL Lac spectral type.
The black hole mass estimated from $\sigma_{e}$ are distributed over a narrower
range because of the relatively small volume sampled ($<z> \sim 0.17$).
}
\end{figure}

\begin{figure}
\plotone{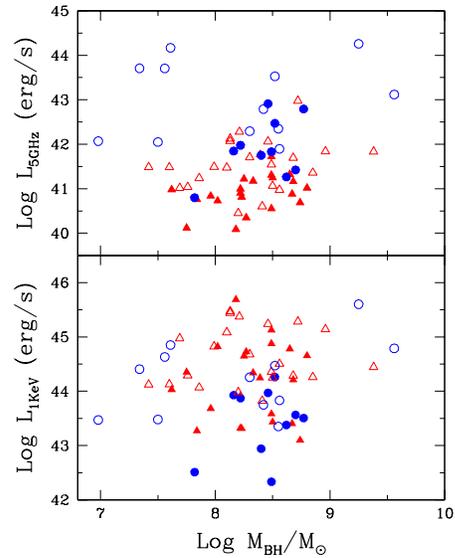}
\caption{Radio and X-ray luminosity vs.~black hole mass.
Neither X-ray nor radio luminosity is correlated with black hole mass.
LBLs tend to have higher radio and lower X-ray luminosity for
a given black hole mass. Symbols are the same as in Figure 3.
}
\end{figure}

\begin{figure}
\plotone{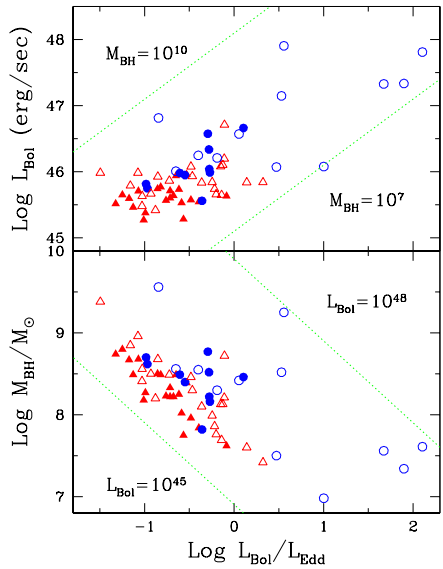}
\caption{
Bolometric luminosity ({\it top}) and
black hole mass ({\it bottom}) vs.~Eddington ratio.
LBLs and HBLs are overlapping in black hole mass. However, LBLs generally
have higher Eddington ratios compared with HBLs.
Symbols are the same as in Figure 3.
The absence of higher and lower black hole mass AGNs in the top panel
as well as higher and lower luminosity AGNs in the bottom panel are likely
caused by selection effects (see text).
}
\end{figure}

\begin{figure}
\plotone{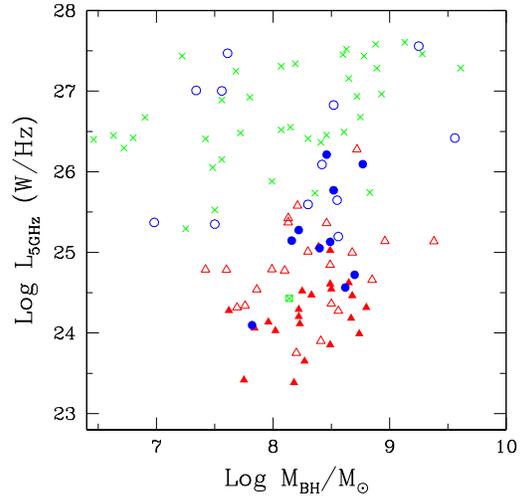}
\caption{Radio luminosity (uncorrected for beaming) of BL Lac objects and FSRQs.
The radio luminosity spans over $\sim 4$ orders of magnitude
across the mass range of $3 \times 10^{7} < M_{\bullet}/M_{\odot} < 10^{9}$.
BL Lac objects generally have low radio luminosity compared with FSRQ.
FSRQs are represented with crosses (colored in
green in electronic edition).
Other symbols are the same as in Figure 3.
A double-peaked emission line source,
PKS 0921-213 (cross with a box), shows much lower radio luminosity compared with other
FSRQs.
%, probably because of its low accretion rate.
}
\end{figure}

\begin{figure}
\plotone{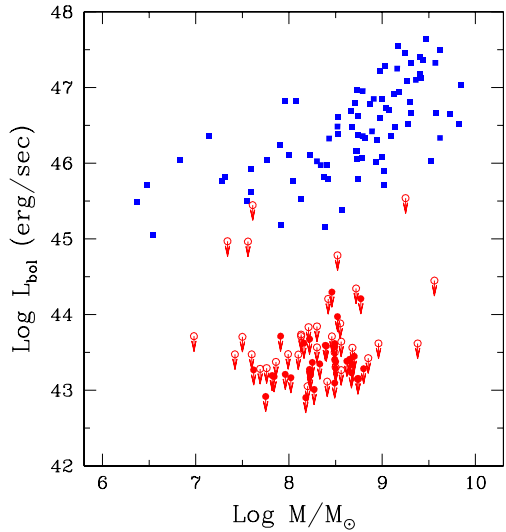}
\caption{Bolometric luminosity vs. black hole mass for radio-loud AGNs.
The bolometric luminosities of BL Lac objects are presented as an upper limit,
after beaming correction with an averaged lower limit to the Doppler factor,
$\delta \gtrsim 3.9$.
For the observed black hole mass range $10^{7} < M_{\bullet}/M_{\odot} < 4 \times 10^{9}$,
the bolometric luminosity spans over 4 orders of magnitude,
indicating a large difference in Eddington ratio among radio-loud AGNs.
{\it Squares:} radio-loud AGN from Woo \& Urry (2002a);
{\it filled circles:} BL Lac objects with black hole mass from the measured $\sigma_{e}$;
{\it open circles:} BL Lac objects with black hole mass from $r_{e}$ and $\mu_{e}$.
}
\end{figure}

\begin{figure}
\epsscale{0.8}
\plotone{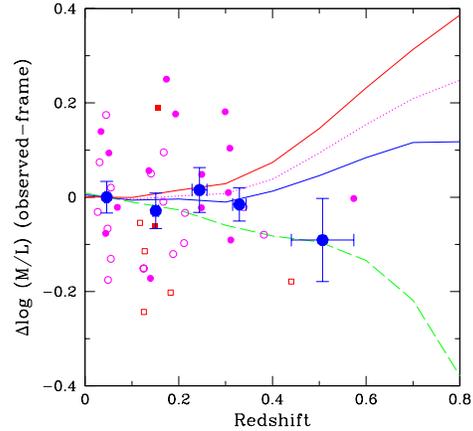}
\epsscale{1.0}
\caption{Evolution of the observed-frame mass-to-light ratio of AGN host galaxies in the Cousins $R$ band.
The evolution of the mass-to-light ratio is consistent with single burst models with $z_{form}=1.4^{+0.9}_{-0.2}$ within the
68\% confidence level. $\Delta \log (M/L)$ is defined as the difference in logarithm
between the $M/L$ at each redshift and the $M/L$ in our lowest AGN host redshift bin
($z=0.046$).
{\it Large filled circles:} averaged $\Delta log (M/L)$ for each redshift bin with 1 $\sigma$ error bars;
{\it small circles:} individual host galaxies of BL Lac objects;
{\it squares:} individual radio galaxies;
{\it small open symbols:} individual host galaxies with mass less than $5 \times 10^{11} M_{\odot}$;
{\it small filled symbols:} individual host galaxies with mass greater than $5 \times 10^{11} M_{\odot}$;
{\it dashed line:} stellar population synthesis model with single burst at $z_{form}=1$;
{\it solid line:} a single burst model with $z_{form}=1.5$;
{\it dotted line:} a single burst model with $z_{form}=2$;
{\it thin-solid line:} a single burst model with $z_{form}=5$.
For clarity no error bars are shown on the measurements for individual galaxies.
For these the reader is referred to Figure~10, which shows the same
measurements with error bars, albeit in the rest-frame rather than
the observed frame.
}
\end{figure}

\begin{figure}
\plotone{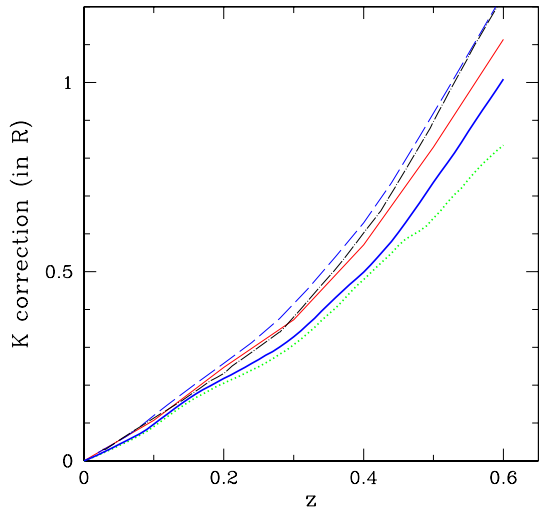}
\caption{$K$-correction values for different stellar population synthesis models.
{\it Dotted line:} a single burst at $z=1$;
{\it solid line:} a single burst at $z=1.4$;
{\it thin-solid line:} a single burst at $z=5$ with a cosmology with $\Omega=0.3$, $\Lambda=0.7$, and $H_{o}=70$ km sec$^{-1}$ Mpc$^{-1}$.
{\it Dashed line:} $K$-correction values from Poggianti 1997, where a cosmology with $q_{o}=0.225$, $H_{o}=50$ km sec$^{-1}$ Mpc$^{-1}$ was used.  
{\it Dot-dashed line:} $K$-correction values from Fukugita et al. 1995, based on the observed
spectra of nearby galaxies with no stellar population evolution using a cosmology with $q_{o}=0$
and $H_{o}=50$ km sec$^{-1}$ Mpc$^{-1}$.
}
\end{figure}

\begin{figure}
\plotone{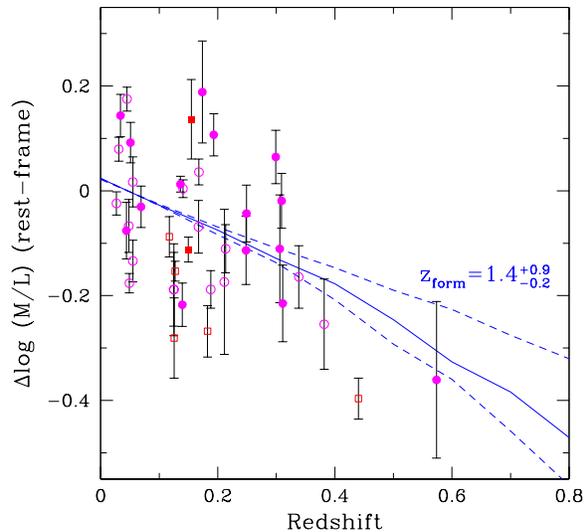}
\caption{Evolution of the rest-frame mass-to-light ratio of AGN host galaxies in the Cousins $R$ band.
The mass-to-light ratio for individual galaxies are derived after $K$-correction with the $z_{form}=1.4$ model.
%Excluding the last z bin, averaged M/L for each redshift bin shows $\Delta log (M/L_{R})/ \Delta z = -0.502$.
{\it Circles:} host galaxies of BL Lac objects;
{\it squares:} radio galaxies;
{\it open sybmols:} galaxies with mass $< 5 \times 10^{11} M_{\odot}$;
{\it filled symbols:} galaxies with mass $> 5 \times 10^{11} M_{\odot}$;
%{\it larger circles:} averaged $\Delta log (M/L)$ for each redshift bin with 1 $\sigma$ error bars;
{\it solid line:} a single burst model with $z_{form}=1.4$, which has
$\Delta log (M/L)/ \Delta z = -0.502$ between $z=0$ and $z=0.4$;
{\it dashed lines:} single burst models with $z_{form}=1.2$ and 2.3, showing 
the uncertainty range determined in the observed mass-to-light ratio evolution (see \S 4.2).
The observed trend is similar to that of early-type galaxies (Treu et al. 2002; van Dokkum \& Franx 2001; Treu et al. 2002).
}
\end{figure}

\begin{figure}
\plotone{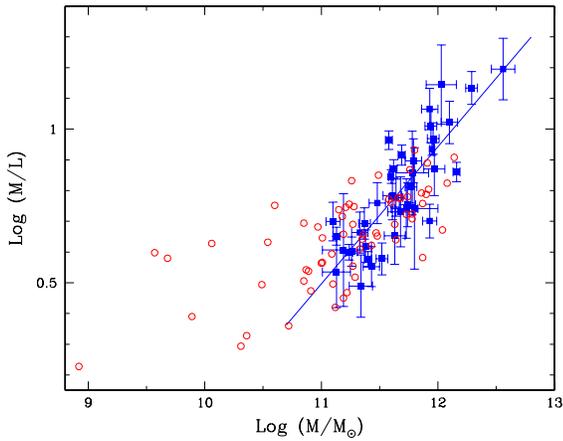}
\caption{Galaxy mass-to-light ratio vs. galaxy mass. The mass-to-light ratios for our AGN host
galaxies ({\it filled squares}), corrected for luminosity evolution using a $z_{form}=1.4$ model,
are compared with those of nearby elliptical galaxies ({\it open circles}) from  van der Marel
\& van Dokkum (2005).
The mass-to-light ratios in the Cousins $R$ band for the nearby galaxies were calculated using
$B-R=1.57$.
The solid line corresponds to the best fit to the AGN host galaxies.
The mass-to-light ratio vs. mass relation of the AGN host galaxies is consistent with that of
normal galaxies in the mass range where the data sets overlap ($11 < log~M/M_{\odot} < 12$).
}
\end{figure}

\clearpage
\begin{deluxetable}{lcrrrrrrrrr}
\rotate {}
\tablewidth{0pt}
\tablecaption{Journal of observations}
\tablehead{
\colhead{Run} &
\colhead{Date} &
\colhead{Telescope} &
\colhead{Instrument} &
\colhead{Grating} &
\colhead{Slit Width}&
\colhead{Resolution} &
\colhead{Plate Scale} &
\colhead{Spatial Scale} &
\colhead{Seeing}  &
\colhead{Sky}     \\
\colhead{}       &
\colhead{}       &
\colhead{}       &
\colhead{}    &
\colhead{lines mm$^{-1}$}           &
\colhead{arcsec}           &
\colhead{\AA}         &
\colhead{\AA pixel$^{-1}$}        &
\colhead{arcsec pixel$^{-1}$} &
\colhead{arcsec}         &  
\colhead{}         \\
\colhead{(1)} &
\colhead{(2)} &
\colhead{(3)} &
\colhead{(4)} &
\colhead{(5)} &
\colhead{(6)} &
\colhead{(7)} &
\colhead{(8)} &
\colhead{(9)} &
\colhead{(10)} &
\colhead{(11)}}  
\tablecolumns{11}
\startdata
1 & 11/27/03   & Magellan 6.5m & B\&C & 600 & 1   & 2.2& 1.56 & 0.25 & 0.7-0.9 & clear  \\
2 & 6/14-17/04 & Magellan 6.5m & B\&C & 600 & 1   & 2.2& 1.56 & 0.25 & 0.7-0.9 & clear  \\
3 & 4/20/04    & Gemini-South  & GMOS & 400 & 0.5 & 1.4& 0.69 & 0.07 & $\sim$ 0.5  & clear  \\
\enddata
\label{observation}
\tablecomments{
Col. (1): observing run.
Col. (2): observing date.
Col. (3): telescope.
Col. (4): instrument.
Col. (5): grating.
Col. (6): slit width.
Col. (7): approximate instrumental resolution in Gaussian $\sigma$.
Col. (8): plate scale.
Col. (9): spatial scale.
Col. (10): seeing FWHM from guiding cameras.
Col. (11): sky condition.
}
\end{deluxetable}

\begin{deluxetable}{lcrrrrrrrrrrrrr}
\rotate {}
\tablewidth{0pt}
\tablecaption{Targets and Measurements}
\tablehead{
\colhead{Name}        &
\colhead{z}           &
\colhead{$<\mu_{e}>$} &
\colhead{$A_R$}       &
\colhead{$K_R$}       &
\colhead{$m_R$}       &
\colhead{$r_e$}       &
\colhead{$\sigma$}    &
\colhead{r}           &
\colhead{C}           &
\colhead{Run}         &
\colhead{Exp.}        &
\colhead{S/N}         \\
\colhead{} &
\colhead{} &
\colhead{} &
\colhead{} &
\colhead{} &
\colhead{} &
\colhead{$^{\prime\prime}$}      &
\colhead{km s$^{-1}$}   &
\colhead{$^{\prime\prime}$}      &
\colhead{}            &
\colhead{ } &
\colhead{hour}        &
\colhead{ } \\                    
\colhead{(1)} &
\colhead{(2)} &
\colhead{(3)} &
\colhead{(4)} &
\colhead{(5)} &
\colhead{(6)} &
\colhead{(7)} &
\colhead{(8)} &
\colhead{(9)} &
\colhead{(10)} &
\colhead{(11)} &
\colhead{(12)} &
\colhead{(13)} }
\tablecolumns{13}
\startdata
0158+001 &  0.2991& 20.13&  0.06& 0.327 &18.27&  1.90$\pm$ 0.10&  273.$\pm$  21.  & 1.25 &  1.08 &1 & 2  &  116 \\
0347-121 &  0.1880& 19.12&  0.12& 0.206 &17.72&  1.25$\pm$ 0.05&  188.$\pm$  10.  & 1.25 &  1.07 &1 & 1.5&  122\\
0506-039 &  0.3059& 19.65&  0.22& 0.337 &18.35&  1.60$\pm$ 0.13&  248.$\pm$  39.  & 1.25 &  1.08 &1 & 1.9&  120\\
1133+161 &  0.5736& 19.70&  0.17& 0.937 &19.83&  1.55$\pm$ 0.23&  212.$\pm$  48.  & 1.25 &  1.10 &2 & 4.3&  65\\
1248-296 &  0.3819& 19.00&  0.20& 0.469 &18.87&  1.10$\pm$ 0.05&  245.$\pm$  32.  & 1.25 &  1.09 &2 & 3.5&  84\\
1255+244 &  0.1407& 19.95&  0.03& 0.153 &16.72&  2.50$\pm$ 0.05&  222.$\pm$   6.  & 1.25 &  1.06 &2 & 1  &  100\\
1534+014 &  0.3110& 19.99&  0.15& 0.345 &18.16&  2.00$\pm$ 0.10&  208.$\pm$  23.  & 1.25 &  1.08 &2 & 3  &  85\\
2326+174 &  0.2134& 19.61&  0.15& 0.230 &17.56&  1.80$\pm$ 0.15&  228.$\pm$  16.  & 1.25 &  1.07 &2 & 2.5&  97\\
0350-371 &  0.1679& 19.35&  0.02& 0.186 &17.08&  1.70$\pm$ 0.07&  276.$\pm$  10.  & 1.25 &  1.06 &2 & 0.8&  70\\
3C135$^{a}$    &  0.1274& 18.99&  0.31& 0.135 &17.05&  1.52$\pm$ 0.01&  197.$\pm$   6.  & 1.25 &  1.05 &1 & 1  & 76 \\
3C424$^{a}$    &  0.1256& 19.57&  0.26& 0.133 &16.44&  2.56$\pm$ 0.04&  171.$\pm$  20.  & 1.25 &  1.05 &2 & 2.8&  51\\
3C348$^{b}$    &  0.1549& 23.39&  0.25& 0.171 &15.64& 22.96$\pm$ 0.90&  212.$\pm$  25.  & 1.25 &  1.06 &2 & 1.6&  39\\
3C306.1$^{a}$  &  0.4403& 18.69&  0.27& 0.581 &19.36&  0.90$\pm$ 0.08&  222.$\pm$  13.  & 0.58 &  1.06 &3 & 2.3& 31 \\
\enddata
\label{data}          
\tablecomments{
Col. (1): AGN name.  
Col. (2): measured redshift
Col. (3): average surface brightness within $r_{e}$ in the Cousins $R$ band calculated
from total host galaxy magnitude (Urry et al. 2000), using Equation (2)
Extinction and $K$-corrected.
Col. (4): foreground extinction correction due to our galaxy from Schlegel et al. (1998).
Col. (5): $K$-correction from our passive evolution model with $z_{form}=1.4$.
Col. (6): observed host galaxy magnitude in the Cousins $R$ band from Urry et al. (2000).
%(7) error of host galaxy magnitude.
Col. (7): half-light radius and error from Urry et al. (2000).
Col. (8): measured stellar velocity dispersion and fitting error of velocity dispersion.
Col. (9): extraction radius in arcseconds.
Col. (10): correction factor for velocity dispersions to a 3.4$^{\prime\prime}$ aperture at the distance of the Coma cluster.
Col. (11): observing run.
Col. (12): total exposure time in hours.
Col. (13): signal-to-noise ratio per pixel, measured at 6000 \AA~ in each combined galaxy spectrum.
The S/N ratio is in the observed spectrum, which consists of AGN and galaxy emission.
Thus, the actual S/N ratio for the galaxy absorption lines is much lower.
}
\tablerefs{
a) galaxy magnitude, $r_{e}$, and $<\mu_{e}>$ are from our $HST$ image analysis (see \S~4.1).
b) galaxy magnitude, $r_{e}$, and $<\mu_{e}>$ are from Roche \& Eales (2000).
}
\end{deluxetable}

\begin{deluxetable}{lcrrrrrr}
%\rotate {}
\tablewidth{0pt}
\tablecaption{Black hole masses and host galaxy luminosities}
\tablehead{
\colhead{Name}        &
\colhead{Type}        &
\colhead{z}           &
\colhead{log ($M_{\bullet}/M_{\odot}$)}        &
%\colhead{$\Delta log (M_{\bullet}/M_{\odot}$)} &
\colhead{log ($M_{G}/M_{\odot}$)}              &
%\colhead{$\Delta (log M_{G}/M_{\odot}$)}       &
\colhead{$M_{R}$}       \\
\colhead{(1)} &
\colhead{(2)} &
\colhead{(3)} &
\colhead{(4)} &
\colhead{(5)} &
\colhead{(6)} }
\tablecolumns{6}
\startdata
  0122+090 &H & 0.3384 & 8.49$\pm$ 0.17&11.61$\pm$ 0.08& -23.02\\
  0145+138 &H & 0.1250 & 7.75$\pm$ 0.27&11.13$\pm$ 0.12& -22.16\\
  0158+001 &H & 0.2991 & 8.65$\pm$ 0.15&11.93$\pm$ 0.07& -23.07\\
  0229+200 &H & 0.1396 & 8.68$\pm$ 0.13&11.93$\pm$ 0.06& -23.76\\
  0331-362 &H & 0.3091 & 8.50$\pm$ 0.15&12.10$\pm$ 0.07& -23.60\\
  0347-121 &H & 0.1880 & 8.02$\pm$ 0.11&11.26$\pm$ 0.05& -22.42\\
  0350-371 &H & 0.1679 & 8.67$\pm$ 0.07&11.69$\pm$ 0.03& -22.66\\
  0506-039 &H & 0.3059 & 8.49$\pm$ 0.30&11.78$\pm$ 0.14& -23.21\\
  3C135    &R & 0.1274 & 8.09$\pm$ 0.06&11.24$\pm$ 0.03& -22.27\\
  0521-365 &L & 0.055$^{a}$ &8.52$\pm$ 0.12&11.38$\pm$ 0.05& -22.50\\
  0525+713 &H & 0.2482 & 8.80$\pm$ 0.22&11.97$\pm$ 0.09& -23.58\\
  0548-322 &H & 0.069$^{a}$ &8.22$\pm$ 0.12&11.77$\pm$ 0.05& -23.00\\
  0706+591 &H & 0.125$^{a}$ &8.25$\pm$ 0.22&11.63$\pm$ 0.09& -23.13\\
  0829+046 &L & 0.1737 & 8.46$\pm$ 0.28&12.03$\pm$ 0.13& -22.95\\
  Mrk421   &H & 0.031$^{a}$ &8.22$\pm$ 0.06&11.62$\pm$ 0.03& -22.44\\
  Mrk180   &H & 0.045$^{a}$ &8.23$\pm$ 0.06&11.58$\pm$ 0.03& -22.12\\
  MS 1133.7+1618 &L & 0.5736 & 8.22$\pm$ 0.43&11.80$\pm$ 0.20& -23.90\\
  1212+078 &L & 0.1363 & 8.70$\pm$ 0.04&11.95$\pm$ 0.02& -23.23\\
  1215+013 &R & 0.1173 & 8.33$\pm$ 0.11&11.37$\pm$ 0.05& -22.36\\
  1215-033 &R & 0.1826 & 7.86$\pm$ 0.14&11.43$\pm$ 0.07& -22.93\\
  1ES 1248-296 &H & 0.3819 & 8.49$\pm$ 0.26&11.68$\pm$ 0.11& -23.37\\
  1255+244 &H & 0.1407 & 8.27$\pm$ 0.06&11.60$\pm$ 0.02& -22.58\\
  1342-016 &R & 0.1498 & 8.47$\pm$ 0.07&12.16$\pm$ 0.03& -23.96\\
  3C306.1  &R & 0.4403 & 8.28$\pm$ 0.11&11.52$\pm$ 0.05& -23.42\\
  1514-241 &L & 0.0490 & 8.40$\pm$ 0.06&11.40$\pm$ 0.03& -22.65\\
  1534+014 &L & 0.3110 & 8.16$\pm$ 0.22&11.73$\pm$ 0.10& -23.38\\
  3C348    &R & 0.1549 & 8.03$\pm$ 0.23&12.56$\pm$ 0.10& -24.13\\
  MRK501   &L & 0.034$^{a}$ &8.62$\pm$ 0.11&11.94$\pm$ 0.05& -22.88\\
  1Zw187   &H & 0.055$^{a}$ &7.84$\pm$ 0.15&11.10$\pm$ 0.06& -21.59\\
  3C371    &L & 0.051$^{a}$ &8.49$\pm$ 0.11&11.96$\pm$ 0.05& -23.05\\
  1853+671 &H & 0.2113 & 7.62$\pm$ 0.41&11.19$\pm$ 0.18& -22.25\\
  1959+650 &H & 0.048$^{a}$ &7.96$\pm$ 0.16&11.33$\pm$ 0.07& -22.24\\
  3C424    &R & 0.1256 & 7.80$\pm$ 0.24&11.34$\pm$ 0.10& -22.79\\
  2143+070 &H & 0.2490 & 8.39$\pm$ 0.16&11.79$\pm$ 0.07& -23.07\\
  2201+044 &L & 0.027$^{a}$ &7.82$\pm$ 0.07&11.13$\pm$ 0.03& -21.76\\
  2254+074 &L & 0.1932 & 8.77$\pm$ 0.12&12.29$\pm$ 0.05& -23.65\\
  2326+174 &H & 0.2134 & 8.33$\pm$ 0.14&11.64$\pm$ 0.06& -22.93\\
  2344+514 &H & 0.044$^{a}$ &8.74$\pm$ 0.18&11.74$\pm$ 0.07& -23.05\\
  2356-309 &H & 0.1671 & 8.18$\pm$ 0.15&11.48$\pm$ 0.07& -22.53\\
\enddata
\label{data2}          
\tablecomments{
Col. (1): AGN name.  
Col. (2): spectral type (L: low frequency peaked BL Lac objects, H: high frequency peaked BL Lac objects, R: radio galaxies),
Col. (3): redshift measured from our observations (measurement errors are typically less than 0.0001).
Col. (4): black hole mass estimated from $\sigma$, using $M_{\bullet} \propto \sigma^{4.02}$ (Tremaine et al. 2002) and
error in black hole mass, derived from $\sigma$ measurement error only.
Col. (5): host galaxy mass from Eq.~\ref{hostM} and error.
Col. (6): absolute R magnitude, extinction and $K$-corrected.
Mass and magnitude are calculated using $H_o=70~km s^{-1} Mpc^{-1}$.
}
\tablerefs{
a) redshift from Urry et al. (2000).
}
\end{deluxetable}

\end{document}